\documentclass[runningheads]{llncs}
\usepackage[utf8]{inputenc}
\usepackage[english]{babel}
\usepackage{graphicx}
\usepackage{xcolor}
\usepackage{amsmath}
\usepackage{amssymb}
\usepackage[normalem]{ulem}
\usepackage[color=yellow!20]{todonotes}
\usepackage{soul}
\usepackage{hyperref}
\usepackage{paralist}
\usepackage{mathrsfs}
\usepackage{listings}
\usepackage{lscape} 
\usepackage{dashbox} 
\usepackage[shortlabels]{enumitem} 

\definecolor{lightgray}{rgb}{0.9,0.9,0.9}
\let\subparagraph\paragraph

 \usepackage{titlesec}
 \titleformat{\paragraph}[runin]
   {\normalfont\normalsize\bfseries}{\theparagraph.}{1em}{}[.]

\usepackage[paperheight=23.5cm,paperwidth=15.5cm,text={12.2cm,19.3cm},centering]{geometry} 

\usepackage[innerleftmargin=5pt,innerrightmargin=5pt]{mdframed}

\usepackage[advantage, asymptotics, adversary, complexity, sets, keys, ff, notions, lambda, primitives, events, operators, probability, logic, mm, landau]{cryptocode}

\usepackage{bm}

\usepackage[capitalise]{cleveref}
\usepackage{caption}
\usepackage{subcaption}
\usepackage[rightcaption]{sidecap}

\usepackage[mathscr]{euscript}

\usepackage[all]{nowidow}

\definecolor{darkred}{rgb}{0.7,0,0}
\definecolor{lightblue}{rgb}{0.85,0.85,1}
\definecolor{bananamania}{rgb}{0.98, 0.91, 0.71}

\newcommand{\comment}[1]{}
\newcommand{\variable}[1]{\mathsf{#1}}

\newcommand{\alg}[1]{\mathsf{#1}}
\newcommand{\smallset}[1] {\{#1\}}
\newcommand{\inp}{\mathsf{pub\_inp}} 
\newcommand{\params}{\variable{pp}} 

\newcommand{\game}[1]{\mathsf{#1}}
\newcommand{\gamestyle}[1]{\mathsf{#1}}
\newcommand{\ufcma}{\gamestyle{UF\mhyphen{}CMA}} 
\newcommand{\cesuf}{\gamestyle{CES\mhyphen{}UF}} 

\newcommand{\curvestyle}[1]{\mathsf{#1}}
\newcommand{\BNCurve}{\curvestyle{BN\mhyphen{}254}}

\newcommand{\algostyle}[1]{\mathsf{#1}}
\newcommand{\groth}{\algostyle{Groth16}}
\newcommand{\oracle}[1]{\algostyle{O}^{#1}} 
\newcommand{\oracleSig}{\oracle{\sig(\sk, \cdot)}} 
\newcommand{\oracleBlsSig}{\oracle{\BLS.\sig(\sk, \cdot)}} 
\newcommand{\oracleCesBlsSig}{\oracle{\BLSCES.\sig(\sk, \cdot)}} 

\newcommand{\paramGen}{\mathcal{G}}
\newcommand{\HashToGOne}{\algostyle{Hash\_to\_G1}}
\newcommand{\GOneToString}{\algostyle{G1\_to\_string}}
\newcommand{\StringToGOne}{\algostyle{string\_to\_G1}}

\newcommand{\Aggr}{\algostyle{Agg}} 
\newcommand{\VerifyAggr}{\algostyle{VfAgg}} 

\newcommand{\signature}{\mathit{\sigma}} 
\newcommand{\msg}{\variable{msg}} 

\newcommand{\CES}{\mathit{CES}} 
\newcommand{\CESSign}{\algostyle{CESSign}}
\newcommand{\CESExtract}{\algostyle{CESExtract}}
\newcommand{\CESVerify}{\algostyle{CESVerify}}
\newcommand{\sigFull}{\mathit{\signature_F}} 
\newcommand{\sigExt}{\mathit{\signature_E}} 

\newcommand{\GRP}{\mathbb{G}} 
\newcommand{\GRPord}{r} 
\newcommand{\pair}{e} 
\newcommand{\ggen}{\mathfrak{g}} 
\newcommand{\hgen}{\mathfrak{h}} 
\newcommand{\isEq}{\stackrel{?}{=}}
\newcommand*{\QED}{\hfill\ensuremath{\square}} 
\newcommand{\cardinality}[1]{\#{#1}}

\newcommand{\role}[1]{\mathrm{#1}}
\newcommand{\CredIssuer}{\role{CredIssuer}}
\newcommand{\CredHolder}{\role{CredHolder}}
\newcommand{\CredVerifier}{\role{CredVerifier}}

\newcommand{\BLS}{\mathtt{BLS}}

\newcommand{\subCred}{\variable{subCred}} 
\newcommand{\cred}{\variable{Cred}} 
\newcommand{\BLSCES}{\mathtt{BLSCES}}
\newcommand{\ZKBLSCES}{\mathtt{ZKBLSCES}}
\newcommand{\SNARK}{\mathtt{SNARK}}
\newcommand{\credSize}{\variable{N}}
\newcommand{\CEAS}{\variable{CEAS}}
\newcommand{\extractionSet}{\variable{X}}

\newcommand{\Extract}{\alg{Extract}}

\newcommand{\Cl}{\alg{ClearIndices}}

\mathchardef\mhyphen="2D

\makeatletter
\renewcommand\subsubsection{\@startsection{subsubsection}{3}{\z@}%
	{-18\p@ \@plus -4\p@ \@minus -4\p@}%
	{0.5em \@plus 0.22em \@minus 0.1em}%
	{\normalfont\normalsize\bfseries}}
\makeatother
\title{A note on anonymous credentials using BLS signatures}
\author{Antoine Rondelet}
\institute{Clearmatics, UK \newline
  \email{ar@clearmatics.com}}
\date{\today}

\begin{document}
\maketitle

\begin{abstract}
    In this note, we remark that the aggregation property of the BLS signature scheme yields an efficient Content Extraction Signature ($\CES$). This construction can be used to build digital credentials that support selective disclosure in various settings. Interestingly, this construction is efficient and well suited to build credential issuance schemes with various applications in the client-server or in the distributed ledger models.
    Finally, we sketch a protocol that combines the $\CES$ with the use of a $\nizk$ which allows to prove predicate satisfaction on claims extracted from a credential, while keeping the data secret.
\end{abstract}

\section{Introduction}

Boneh, Lynn and Shacham (BLS)~\cite{DBLP:conf/asiacrypt/BonehLS01} introduced a digital signature scheme leveraging pairing based cryptography. These ``BLS signatures" are constructed using Gap Diffie-Hellman groups, in which the Decisional co-Diffie-Hellman (co-DDH) problem is easy but the Computational co-Diffie-Hellman (co-CDH) problem remains hard~\cite{DBLP:conf/asiacrypt/BonehLS01}.
The BLS signature algorithm is deterministic\footnote{It is well known that randomness re-use is a deadly mistake (see e.g.~\cite{DBLP:journals/iacr/BreitnerH19,schnorrBiasedNonce}), which makes deterministic algorithms appealing.}, it supports signature aggregation, it can easily be used to define multi-signatures protocols and produces very short signatures. Due to all these properties, the BLS signature scheme has gained traction over the past years. It is now undergoing standardization\footnote{\url{https://tools.ietf.org/html/draft-boneh-bls-signature-02}} and is of great interest for various blockchain protocols, e.g.~\cite{DBLP:conf/asiacrypt/BonehDN18,celo-docs}.

While the aggregation property of BLS signatures~\cite{DBLP:conf/eurocrypt/BonehGLS03} is well-known and appreciated, we show how this property can be used to build a Content Extraction Signature ($\CES$).
Introduced by Steinfeld et al.~\cite{DBLP:conf/icisc/SteinfeldBZ01} Content Extraction Signatures are digital signature schemes enabling users to generate signatures on an extracted portion of a signed document. In their original paper~\cite{DBLP:conf/icisc/SteinfeldBZ01} the authors initially presented $\CES$ built from commitment schemes or from a digital signature scheme based on RSA~\cite{DBLP:conf/eurocrypt/BellareGR98}.

We noticed that the ability to aggregate BLS signatures can be used to obtain a Content Extraction Signature scheme. This scheme can be used to build digital credentials, and integrates well with blockchain projects such as Ethereum~\cite{ethereum-whitepaper} which introduced precompiled contracts to carry out elliptic curve arithmetic and pairing checks as part of the state transition function\footnote{We note that, due to improvement on the Number Field Sieve (NFS) algorithm to solve the discrete logarithm problem in finite fields~\cite{bnNFS}, the $\BNCurve$ precompiled contracts, introduced in the Byzantium release of the Ethereum protocol, now fail to address the security level initially targeted (128 bits of security). New precompiled contracts, however, are expected to be introduced to do arithmetic on other curves~\cite{eip1962}}.

Before presenting our work, it is important to note that our observation - to use the BLS signature as a $\CES$ - had been independently noticed several years ago by the authors of~\cite{DBLP:conf/icisc/SteinfeldBZ01}. In fact, this idea was briefly mentioned as a note in a subsequent version of the full document we received after request to the authors for more details on their work~\cite{DBLP:conf/icisc/SteinfeldBZ01}.
Nevertheless, and to the best of our knowledge, no explicit description of the $\CES$ based on BLS signature was presented. We aim to fill this gap in this note as we believe this constitutes another interesting use case for the BLS signature scheme. Our study will be centered around anonymous credential use-cases and will heavily be based on~\cite{DBLP:conf/icisc/SteinfeldBZ01}. As such, we assume familiarity with this paper. We will conclude this note by sketching how this BLS-based $\CES$ can be coupled with efficient $\nizk$s in order to go ``beyond selective disclosure'', and remove the need to disclose any sensitive information related to a user's credentials.

\section{Background and Notations}

In the rest of the document, let $\ppt$ denote probabilistic polynomial time.
Let $\secpar \in \NN$ be the security parameter.
We assume all algorithms described receive as an implicit parameter the security parameter written in unary $\secparam$.
All protocol participants are modeled as probabilistic Turing Machines.

In the following sections, we will denote by $||$ the concatenation of two strings, i.e.~$|| \colon \{0, 1\}^n \times \{0, 1\}^m \to \{0, 1\}^{m+n}$. The length of a string $s$ is given by $|s|$, and the cardinality of a set $S$ will be given by $\cardinality{S}$. We denote by $[n]$ the set $\{0, \ldots, n-1\}$ for $n \in \NN$.

An adversary and all parties they control are denoted by $\adv$.
We write $\negl[]$ or $\poly[]$ to denote a negligible or polynomial function respectively.
We say that an adversary wins a game $\game{GAME}$ if they make $\game{GAME}$ return $1$.
We say that $\adv$ wins $\game{GAME}$ with negligible advantage if $\advantage{\game{GAME}}{\adv} = \abs{\prob{\game{GAME} = 1} - 1/2} \leq \negl$.

\subsection{Digital signatures}\label{subsec:digSig}

Informally, a digital signature scheme consists of 3 algorithms:
\begin{itemize}
    \item $\kgen$ (Key generation algorithm): Takes the security parameter $\secparam$ as input and outputs a key pair $(\sk, \pk)$.
    \item $\sig$ (Signing algorithm): Takes the signer's $\sk$ and the message to sign $\msg$ as input and returns a signature $\signature$.
    \item $\verify$ (Verification algorithm): Takes the signature $\signature$, the message $\msg$ and the signer's public key $\pk$ as inputs, and returns 1 if $\signature$ is a valid signature, else 0.
\end{itemize}

For a signature scheme to be secure, it needs to be unforgeable (i.e.~it should be intractable to produce a signature, without knowing the signing key $\sk$, on a message that has not been signed yet). One approach to prove a digital signature scheme to be secure, is to use a game-based proof in which the probability for a $\ppt$ adversary $\adv$ to win the ``unforgeability under adaptive chosen message attack'' game, that we denote by $\ufcma$ (see~\cref{fig:uf-cma-game}), is negligible in the security parameter (i.e.~$\negl[\secparam]$). In this game, the adversary can query a signing oracle - $\oracleSig$ (having the signing key) - on a set of messages, in order to obtain the corresponding set of signatures. However, $\adv$ should not be able to leverage this set of message/signature pairs in order to forge a signature on a message that was never queried to the oracle.

\begin{figure}
    \begin{minipage}[t]{\textwidth}
        \centering
        \procedure[linenumbering]{$\ufcma(\secparam, q)$}{%
            (\sk, \pk) \gets~\kgen(\secparam) \\
            \state \gets~\adv^{\oracleSig}(\pk,\cdot) \\
            \pccomment{$\state = \{(\msg_i, \sigma_i)\}_{i \in [q]}$ where $\msg_i$ denotes the ith query} \\
            \pccomment{made to $\oracleSig$ and $\sigma_i$ denotes the ith oracle answer} \\
            (\msg^{*}, \sigma^{*}) \gets~\adv(\pk, \state) \\
            \pcreturn \verify(\pk, \msg^{*}, \sigma^{*}) = 1 \land \msg^{*} \not \in \{\msg_i\}_{i \in [q]}
        }
        \caption{$\ufcma$ game}
        \label{fig:uf-cma-game}
    \end{minipage}%
\end{figure}

\subsection{Design goals of Content Extraction Signatures}

In their work~\cite{DBLP:conf/icisc/SteinfeldBZ01} Steinfeld et al.~introduced the notion of Content Extraction Signatures.

These signatures enable a user to extract part of a signed document (where a document is modeled as a set of messages indexed by the set of natural numbers $\NN$) to produce a signed sub-document (containing a subset of the initial document's messages). This sub-document can then be presented to a verifier who can verify that all messages have been signed by the signer. Doing so enables to keep a set of messages (potentially sensitive pieces of information) secret and undisclosed\footnote{We will use set operations on documents in the rest of this note. For instance, let $D$ and $D'$ be two documents, then if $D' \cap D = \emptyset$, this means that $D'$ and $D$ have no message in common. Likewise, we will write $S \subset D$ to denote that $S$ is a sub-document of $D$, $\cardinality{D}$ returns the number of messages in $D$ etc.}.

Using a $\CES$ presents two advantages. First, it lessens the tension between information granularity and user's information privacy. Additionally, it balances information granularity and bandwidth usage.
We see that, a $\CES$ enables users to embrace a \emph{minimal information model}, in which they disclose the bare minimum of the information needed to undertake an action.

In order to help the reader fully appreciate their contributions, the authors of~\cite{DBLP:conf/icisc/SteinfeldBZ01} propose a comparison of their $\CES$ constructions with a ``simplistic approach''. In this naive approach, the signer signs each individual message of a document (denoted $D = \{\msg_1, \ldots, \msg_n\}$) independently and sends the set of signed messages, $\widetilde{D} = \{(\msg_1, \signature_1), \ldots, (\msg_n, \signature_n)\}$, to the document holder. After receiving this set of signed messages, the holder is free to send the desired signed sub-document - $SD = \{(\msg_1, \signature_1), \ldots, (\msg_m, \signature_m)\}$, $m \leq n$ - to the verifier in order to authenticate.
It is quite easy to see that such a simplified approach is everything but optimal. In fact, when presenting a sub-document of $N$ messages to the verifier, the holder now needs to adjoin the $N$ corresponding signatures. Hence, the size  of the signature for the disclosed messages grows linearly with the number of messages extracted from the initial document. Likewise, the routine described above compels the signer to call the $\sig$ algorithm for each message in the document $D$. The computational and bandwidth overhead of this ``simplistic approach'' are clear and not desired.

The $\CES$ constructions provided in~\cite{DBLP:conf/icisc/SteinfeldBZ01} address either the computational or bandwidth issues above-listed.
Last but not least, splitting the document into messages and sending the individual messages to a signer \emph{does not provide any guarantee about the respect of the semantics of the initial document}. Some messages in the document $D$ can be related, and so, ensuring an ordering relation on the messages composing a document would be highly desirable. This would prevent changing the semantics of the document by presenting a re-ordered set of signed messages to the verifier\footnote{re-ordering a list of facts might change the initial overall meaning.}. While such integrity insurance is clearly not provided by the naive approach detailed above, this has been addressed by using a \emph{Content Extraction Access Structure} (CEAS) as part of the $\CES$ construction. We refer the reader to~\cite{DBLP:conf/icisc/SteinfeldBZ01} for more details.

\paragraph{Overview of $\CES$}

A $\CES$ comes with various requirements. We briefly remind these requirements below but, again, gently advise the interested reader to refer to~\cite{DBLP:conf/icisc/SteinfeldBZ01} for more context and formal definitions.

Informally, a $\CES$ needs to ensure the following requirements:
\begin{itemize}
    \item Extraction: Given a document, it should be possible for anyone to extract a sub-document associated with a valid and publicly verifiable signature.
    \item Efficiency: A $\CES$ should be more efficient (communication-wise \emph{or} computationally) than the ``simplistic approach'' aforementioned.
    \item Iterative Extraction: It should be possible to extract valid sub-documents and associated extracted signature from a sub-document (itself extracted from the source/initial document).
\end{itemize}

Now that we have introduced the requirements for a $\CES$ scheme, we provide a definition of such signatures.
A $\CES$ is defined as a set of algorithms:
\begin{itemize}
    \item $\kgen$ (Key generation algorithm): Generates $(\sk, \pk)$ on input $\secparam$.
    \item $\CESSign$ (Signature algorithm): Takes $\sk$, a document $D$ and a CEAS\footnote{The CEAS can be seen as a specification of the ``extraction rules'' (e.g.~``to be valid, an extracted signature should be associated with a sub-document in which sub-message $i$ is non-blinded''). The CEAS allows the signer to define some extraction rules that need to be followed for the extracted signature to be valid. In other words, the CEAS defines the class of valid extracted signatures from a document.} as inputs, and outputs a signature $\sigFull$ for the document $D$.
    \item $\CESExtract$ (Extraction algorithm): Takes $\pk$, a document $D$, a signature $\sigFull$, and an extraction subset (specifies the extraction to be carried out, i.e.~the indices of each messages of the document to add to the sub-document) as inputs, and outputs an extracted signature $\sigExt$.
    \item $\CESVerify$ (Verification algorithm): Takes $\pk$, an extracted sub-document, and an extracted signature $\sigExt$ as inputs and outputs 1 or 0 for acceptance or rejection of the extracted signature.
\end{itemize}

\paragraph{A note on security}

The attentive reader would remark that the standard notion of ``unforgeability''~\cref{subsec:digSig}, which is central to argue about the security of signature schemes, is trivially broken in the context of a $\CES$. In other words, the definition of a $\CES$ provides the adversary $\adv$ with a way to win the $\ufcma$ game with non negligible probability. In fact, in this setting, any proper sub-document of the signed document is considered as a new message and therefore the extraction of a valid signature for it constitutes an existential forgery (in the traditional sense). Hence, the model needs to be relaxed to allow signatures on extracted documents (that follow the CEAS) to be acceptable by the verification algorithm. This idea is represented by the $\CES$-Unforgeability requirement introduced in~\cite{DBLP:conf/icisc/SteinfeldBZ01}.

In addition to CES-Unforgeability (denoted $\cesuf$), a $\CES$ needs to prevent leakages about undisclosed messages of a document (blinded in the extracted sub-document). This is captured by the $\CES$-Privacy requirement in~\cite{DBLP:conf/icisc/SteinfeldBZ01}.

\subsection{Credentials data model}\label{subsec:cred-model}

In this section we consider every documents to be credentials. We base our model on definitions taken from~\cite{vcModelw3c}\footnote{``W3C Recommendation 19 November 2019'' is the latest version of the document at the time of writing.}.

\paragraph{Claim}
A claim is an assertion made about a subject. We follow~\cite[Section 3]{vcModelw3c} and express each claim using \emph{subject-property-value} relationships (e.g.~\emph{Alice-Age-19}). Additionally, we denote by \emph{subject-property} the prefix of the claim. Moreover, if the ``subject'' of the claim is clear from the context, we will remove it from the prefix of the claim.

In addition to the standard definition of a claim, we introduce two types of claims: \emph{visible} claims and \emph{hidden} claims.
We say that a claim is \emph{visible} if \emph{value} is not \emph{BLINDED} in the claim relationship. On the other hand, we say that a claim is \emph{hidden} if \emph{value} is set to \emph{BLINDED} in the claim relationship. In other words, an \emph{hidden claim} is of the form \emph{subject-property-BLINDED}.

\paragraph{Credential}
Roughly speaking, a credential for a subject $S$ is a set of claims made by an issuer about $S$.

\paragraph{Verifiable Presentation and Extracted credentials}

By default, all claims are \emph{visible} when the credential is issued by the issuer. We describe below how the credential holder can derive a presentation that only discloses the desired information.

According to~\cite{vcModelw3c}, a presentation is the ``data derived from one or more verifiable credentials, issued by one or more issuers, that is shared with a specific verifier''. The notion of ``verifiability'' is bound to the notion of cryptographic verification which aims to detect credential authorship tampering with overwhelming probability.
The objective behind ``verifiable presentations'' is to make sure that authorship of the data is not altered after the adversary transforms an issued credential into a presentation.

Below, we describe the process of extracting ``sub-credentials'' from an issued credential.

Let $\subCred$ represent a \emph{sub-credential} of an initial credential $\cred$. Informally, we say that $\subCred$ is extracted from $\cred$ if all the visible claims in $\subCred$ are in the set of (visible) claims contained in $\cred$\footnote{We remind that all claims are visible by default in an issued credential.}. More precisely, we define $\subCred$ --- a sub-credential of $\cred$ --- as a set of claims such that:
\begin{itemize}
    \item At least one claim in $\subCred$ is hidden
    \item All visible claims in $\subCred$ are in $\cred$, and have the same index
    \item Both $\subCred$ and $\cred$ have the same number of claims, i.e.~$\cardinality{\cred} = \cardinality{\subCred}$
\end{itemize}

\paragraph{Roles}

In this document, we follow the roles introduced in~\cite{vcModelw3c}.
\begin{itemize}
    \item $\CredIssuer$: Issues the credentials to the credential holder $\CredHolder$.
    \item $\CredHolder$: Receives credentials from a set of $\CredIssuer$s, stores the credentials and presents them to a set of verifiers ($\CredVerifier$) to authenticate to various systems. While the credential holder may not necessarily be the credential subject (entity about which claims are made) in all circumstances\footnote{E.g.~ The subject of a credential can be a pet, say a dog, while the credential holder can be the pet owner.}, we assume without loss of generality, that the holder is the subject.
    \item $\CredVerifier$: Receives and verifies credentials in order to grant or refuse access to some resources.
\end{itemize}

\section{Using the BLS signature as a $\CES$}

In their paper~\cite{DBLP:conf/icisc/SteinfeldBZ01} the authors proposed several constructions for $\CES$. These constructions leverage two main cryptographic tools: commitment schemes and a $\ufcma$ digital signature scheme based on RSA.

In the rest of this section, we show that using the BLS signature along with the credential data model introduced in~\cref{subsec:cred-model} naturally leads to a $\CES$.

\subsection{BLS signature}\label{subsec:bls}

Let $\paramGen$ be a bilinear group generator taking $\secparam$ as input and returning a bilinear group $\params = (\GRPord, \GRP_1, \GRP_2, \GRP_T, \pair)$, where $\GRP_1, \GRP_2, \GRP_T$ are cyclic groups of order $\GRPord$, and $\pair$ is a bilinear map $\pair \colon \GRP_1 \times \GRP_2 \to \GRP_T$, such that $\pair$ is non-degenerate and bilinear. We further assume that arithmetic in the groups and computing $\pair$ is efficient. We denote by $\ggen_1, \ggen_2, \ggen_T$ the generators of the groups $\GRP_1, \GRP_2, \GRP_T$ respectively. We denote by $+$ the group operation in the two source groups $\GRP_1,\GRP_2$. Additionally, we define the infix operator $\cdot \colon \FF_r \times \GRP_{1,2} \to \GRP_{1,2}$ that represents the successive application of the group operation - e.g.~$k \cdot \ggen_1 = \ggen_1 + \ldots + \ggen_1$ ($k$ times). The group operation for the target group $\GRP_T$ is denoted by $*$, and the infix operator $\hat{} \colon \GRP_T \times \FF_r$ denotes its successive application - e.g.~${\ggen_T}^k = \ggen_T * \ldots * \ggen_T$ ($k$ times).

The BLS signature scheme is defined by the three algorithms below:\footnote{Note that we voluntarily place signatures in $\GRP_1$ to minimize the signature size, and limit the credentials size. Another variant of BLS signatures consists in placing public keys in $\GRP_1$ to minimize their size.}

\begin{figure}
    \begin{pchstack}[center]
        \procedure[linenumbering, syntaxhighlight=auto]{$\BLS.\kgen(\secparam)$}{%
            \sk \gets \ZZ_{\GRPord}^{*} \\
            \pk \gets \sk \cdot \ggen_2 \\
            \pcreturn (\sk, \pk)
        }

        \pchspace

        \procedure[linenumbering, syntaxhighlight=auto]{$\BLS.\sig(\sk, \msg)$}{%
            \hgen \gets \HashToGOne(\msg) \\
            \sigma \gets \GOneToString(\sk \cdot \hgen) \\
            \pcreturn (\sigma)
        }

        \pchspace

        \procedure[linenumbering, syntaxhighlight=auto]{$\BLS.\verify(\pk, \msg, \sigma)$}{%
            \hgen \gets \HashToGOne(\msg) \\
            c \gets \pair(\hgen, \pk) \\
            c' \gets \pair(\StringToGOne(\sigma), \ggen_2) \\
            \pcreturn (c \isEq c')
        }
    \end{pchstack}
    \caption{BLS signature algorithms}
\end{figure}

Where $\HashToGOne \colon \{1, 0\}^* \to \GRP_1$ is a cryptographic hash function --- treated as a Random Oracle --- that hashes a string into a group element in $\GRP_1$, $\GOneToString \colon \GRP_1 \to \{1, 0\}^*$ is a function that encodes an element of $\GRP_1$ to a string, and where $\StringToGOne \colon \{1, 0\}^* \to \GRP_1$ is the inverse function of $\GOneToString$ that decodes a string-encoded element of $\GRP_1$ to retrieve the group representation, i.e. $\forall x \in \GRP_1, \StringToGOne(\GOneToString(x)) = x$.

In addition to the three algorithms $(\kgen, \sig, \verify)$, common to every digital signature scheme (see~\cref{subsec:digSig}), the BLS signature supports signature aggregation. To put it simply, the $\Aggr$ algorithm takes $n$ signatures as input and compresses them into a single signature. The $\Aggr$ and $\VerifyAggr$ algorithms are recalled below:

\begin{figure}[h]
    \begin{pchstack}[center]
        \procedure[linenumbering, syntaxhighlight=auto]{$\BLS.\Aggr(\smallset{\sigma_i}_{i=0}^{n-1})$}{%
            \sigma_{\Aggr} \gets \ggen_1 \\
            \pcfor i \in [n] \pcdo \\
            \pcind \sigma_i \gets \StringToGOne(\sigma_i) \\
            \pcind \sigma_{\Aggr} \gets \sigma_{\Aggr} + \sigma_i \\
            \pcendfor \\
            \sigma_{\Aggr} \gets \GOneToString(\sigma_{\Aggr}) \\
            \pcreturn \sigma_{\Aggr}
        }

        \pchspace

        \procedure[linenumbering, syntaxhighlight=auto]{$\BLS.\VerifyAggr(\smallset{\pk_i}_{i=0}^{n-1}, \smallset{\msg_i}_{i=0}^{n-1}, \sigma_{\Aggr})$} {%
            c \gets \prod_{i=0}^{n-1} \pair(\HashToGOne(\msg_i), \pk_i) \\
            c' \gets \pair(\StringToGOne(\sigma_{\Aggr}), \ggen_2) \\
            \pcreturn (c \isEq c') \\
        }
    \end{pchstack}
    \caption{Aggregation of BLS signatures}
\end{figure}

\subsection{BLS signature as $\CES$}

\begin{figure}[h]
    \centering
    \includegraphics[width=1\textwidth]{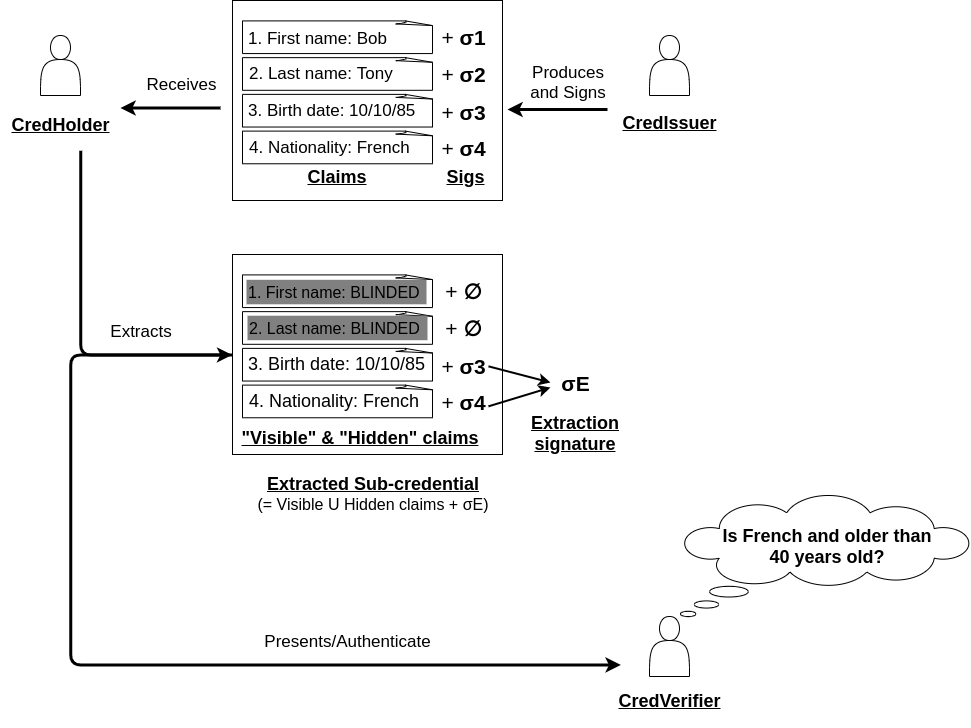}
    \caption{An overview of the content extraction protocol}
    \label{fig:ces-flow}
\end{figure}

In this section we present the BLS-variant of the RSAProd (RSAP) scheme presented in~\cite{DBLP:conf/icisc/SteinfeldBZ01}. An example of how to use the BLS signature as a CES is provided~\cref{fig:ces-flow}.

We denote by $\cred$ a credential (ordered set of claims) of size $\cardinality{\cred} = \credSize$. The extraction set (set of indices of visible claims to extract from a credential $\cred$) is denoted by $\extractionSet$, $\cred[i]$ represents the $i^{th}$ claim of the credential, and $\Cl()$ takes a credential as input and returns the set of indices of the visible claims in the credential.

\newcommand{\signedCred}{\variable{SignedCred}}
\newcommand{\hiddenCred}{\variable{HiddenCred}}
\begin{figure}[h]
\begin{minipage}{.5\textwidth}
    \begin{pcvstack}[center]
        \procedure[syntaxhighlight=auto]{$\BLSCES.\kgen(\secparam)$}{%
            \pcreturn \BLS.\kgen(\secparam)
        }

        \pcvspace

        \procedure[syntaxhighlight=auto]{$\BLSCES.\Extract(\signedCred, \extractionSet)$}{%
            \Sigma \gets \emptyset \\
            \pccomment{$\hiddenCred$ contains all claims in}\\
            \pccomment{$\signedCred.\cred$ but they are all}\\
            \pccomment{hidden by default}\\
            \subCred \gets \hiddenCred \\
            \pcfor i \in \extractionSet \pcdo \\
            \pcind \Sigma \gets \Sigma \cup \signedCred.\sigFull.\sigma_i \\
            \pcind \pccomment{Make the claim visible}\\
            \pcind \subCred[i] \gets \cred[i]\\
            \pcendfor \\
            \sigma \gets \BLS.\Aggr(\Sigma) \\
            \sigExt \gets (\signedCred.\CEAS, \sigma) \\
            \pcreturn (\subCred, \sigExt)
        }
    \end{pcvstack}
\end{minipage}%
\begin{minipage}{.5\textwidth}
    \begin{pcvstack}[center]
        \procedure[syntaxhighlight=auto]{$\BLSCES.\sig(\sk, \cred, \CEAS$)}{%
            \credSize \gets \cardinality{\cred} \\
            \pcfor i \in [\credSize] \pcdo \\
            \pcind \sigma_i \gets \BLS.\sig(\sk, \CEAS || \credSize || i || \cred[i]) \\
            \pcendfor \\
            \sigFull \gets (\CEAS, \smallset{\sigma_i}_{i \in [\credSize]}) \\
            \signedCred \gets (\cred, \sigFull)\\
            \pcreturn \signedCred
        }

        \pcvspace

        \procedure[syntaxhighlight=auto]{$\BLSCES.\verify(\pk, \subCred, \sigExt$)}{%
            \extractionSet' \gets \Cl(\subCred) \\
            \credSize' \gets \cardinality{\subCred} \\
            M \gets \emptyset \\
            \pcfor i \in \extractionSet' \pcdo \\
            \pcind M \gets M \cup \CEAS || \credSize' || i || \cred'[i] \\
            \pcendfor \\
            b \gets (\extractionSet' \in \CEAS) \land \BLS.\VerifyAggr(\pk, M, \sigExt.\sigma)\\
            \pcreturn b
        }
    \end{pcvstack}
\end{minipage}
    \caption{Using the BLS signature as a CES}
    \label{fig:bls-ces}
\end{figure}

We observe that using the BLS signature as above yields a $\CES$ that is $\CES$-Private and $\cesuf$.

\paragraph{CES-Privacy}

We see that the BLS signature trivially yields a $\CES$-Private Content Extraction Signature.

\begin{proof}
No information related to \emph{hidden} claims is used to generate the signature. In fact, the extracted signature $\sigExt$ is obtained as the sum (i.e.~aggregation) of the signatures of the visible claims only.
The BLS trivially yields a $\CES$-Private Content Extraction Signature
\QED
\end{proof}

\paragraph{CES-Unforgeability}

We recall the notion of CES-Unforgeability~\cite{DBLP:conf/icisc/SteinfeldBZ01} and adapt it to fit with our data model.
In order to be coined a $\CES$, it should be infeasible for an attacker $\adv$, having access to a $\CES$ signing oracle, to produce a credential/signature pair $(\cred^*, \signature^*)$, such that:
\begin{enumerate}
    \item $\signature^*$ passes the verification test for $\cred^*$, and
    \item $\cred^*$ is either:
        \begin{enumerate}[(a)]
            \item \label{itm:a} Not a sub-credential of any credential queried to the $\CES$ signing oracle, or
            \item \label{itm:b} Is a sub-credential of a queried credential $\cred$, but not allowed to be extracted by the CEAS attached to $\cred$ in the signing query.
        \end{enumerate}
\end{enumerate}

The BLS signature induces a CES signature which is CES-Unforgeable.

\newcommand{\cclaim}{\variable{claim}}
\begin{proof}
We start by proving by contradiction that~\cref{itm:a} only holds with negligible probability.
To do so, we start by assuming that BLS is $\ufcma$, and thus, we assume that $\adv$ cannot forge BLS signatures. As such:
        \begin{equation}\label{assumption}
            \condprob{\BLS.\verify(\pk, \signature^*, \msg^*) = 1}{
                \begin{aligned}
                    (\sk, \pk) \gets \BLS.\kgen(\secparam)\\
                    \{(\msg_i, \sigma_i)\}_{i \in [q]} \gets \adv^{\oracleBlsSig}\\
                    (\msg^*, \signature^*) \gets \adv(\{(\msg_i, \sigma_i)\}_{i \in [q]})
                \end{aligned}
            } \leq \negl
        \end{equation}

        We now assume that $\adv$ obtained a forgery for the $\CES$ by crafting a message $\cred^*$ that is not a sub-credential of any credential queried to the $\CES$ signing oracle. As such, we have\footnote{Here we assume that $\cclaim$ wasn't previously queried to the BLS signing oracle by $\adv$. As such, we treat the tuple $(\cclaim, \widetilde{\sigma})$ as a BLS forgery.}:

        \[
            \begin{split}
                &[\{(\cred'_i, \sigma_i)\}_{i \in [q']} \gets \adv^{\oracleCesBlsSig}, (\cred^*, \signature^*) \gets \adv(\{(\cred'_i, \sigma_i)\}_{i \in [q']})]\ \land\\
                &\forall\ i \in [q'],\ (\cred^* \not\subseteq \cred'_i)\ \land\ \BLSCES.\verify(\pk, \cred^*, \signature^*) = 1 \\
                &\Rightarrow \exists\ \cclaim \in \cred^* \setminus (\bigcup_{i=0}^{q'-1} \cred^* \cap \cred'_i), (\cclaim,   \widetilde{\sigma}) \gets \adv\ \land\ \BLS.\verify(\pk, \cclaim, \widetilde{\sigma}) = 1\\
                &\Rightarrow \neg \cref{assumption}
            \end{split}
        \]

    Now, we show that~\cref{itm:b} can only hold with negligible probability.
    We assume that $\adv$ finds a pair $(\cred^*, \signature^*)$ such that:
    \[
        \BLSCES.\verify(\pk, \cred^*, \signature^*) = 1\ \land\ \exists i \in [q'], \cred^* \subset \cred'_i
    \]
    is a sub-credential of a queried credential $\cred$ to the signing oracle.
    Additionally, we assume that~\cref{itm:b} holds and that $\cred^*$ is not included in the allowed extraction subsets (i.e.~violates the extraction policy specified in the $\CEAS$). By looking at~\cref{fig:bls-ces} we see that one of the checks carried out as part of the $\BLSCES.\verify$ algorithm consists is verifying that all the visible claims are conform to the  $\CEAS$. As such, the only way for $\adv$ to succeed in his attack is to modify the $\CEAS$ to his advantage. Nevertheless, since the $\CEAS$ is part of the message passed to the $\HashToGOne$ hash function, the only way for $\adv$ to succeed is to find a collision in the hash function. However, this is assumed to be intractable.

    As such, we conclude that BLS fulfills the $\CES$-Unforgeability requirement as initially stated in~\cite{DBLP:conf/icisc/SteinfeldBZ01}
    \QED
\end{proof}

\begin{remark}
    The hidden claims are not manipulated during the verification process. Hence an attacker could maliciously change the claim-prefix of an hidden claim. This attack could trick the verifier in believing that $\adv$ was issued a credential which contains a claim with a given property. Nevertheless, we consider this attack to be minor since an hidden-claim with a maliciously crafted prefix will never be able to become visible.
    In fact, we assume that it would be naive for a $\CredVerifier$ to take any decision based on an hidden claim.

    To avoid the aforementioned scenario, one could also assume that an hidden claim can be such that \textbf{both} the property and value of the claim are set to \emph{BLINDED}. This would avoid potential information leakage about the properties of the hidden claim while protecting against the minor attack vector mentioned above.
\end{remark}

\newcommand{\rhs}{\variable{RHS}}
\section{From selective disclosure to zero knowledge}

While leveraging the aggregation property of the BLS signature naturally translates into a Content Extraction Signature, and thus enabling selective claim disclosure on credentials, many blockchain use-cases may require to send the verifiable presentation of the credential directly on-chain.
In this setting, it is clear that providing selective disclosure may not be enough. In fact, recent data protection regulations like GDPR\footnote{\url{https://eugdpr.org/}} include the right for a subject to request the erasure of his personal data on a given information system. However, it is well known that blockchain systems provide a source of immutability, and, as such, erasing data becomes impossible\footnote{Well, only a fork could be used to ``remove'' some data from the chain.}.
As a consequence, we propose, in this section, a way to acquire stronger privacy guarantees. We propose to use the BLS-CES described above along with an efficient $\nizk$ such as a zk-SNARK, in order to go from ``selective claim disclosure'' to ``zero-knowledge proofs on claims of a credential''. In the following, we will consider that the SNARK-scheme used is~\cite{DBLP:conf/eurocrypt/Groth16}. As such, we assume that the statement to be proven is represented as a Rank-1 Constraint System (R1CS). When referring to computation that is done ``in the SNARK'' we mean ``represented as an R1CS program''.

\subsection{The protocol}

In this section, we assume that upon successful verification of the BLS-CES signature, $\CredVerifier$ is interested in verifying if the disclosed claims satisfy some NP predicate $P$.

As such, the behavior of $\CredVerifier$ can be modeled as the predicate $CV(\pk, \signature, \cred) = \BLSCES.\verify(\pk, \cred, \signature)\ \land\ P(\cred)$. From this, we clearly see that the checks carried out by the $\CredVerifier$, and represented as the predicate $CV$\footnote{Which is an NP predicate as it is a conjunction of NP predicates.}, can all be carried out ``in a SNARK''.
As a consequence, it is possible to extend the computation carried out by $\CredHolder$ in order to generate a SNARK, $\pi$, for the predicate $CV$. The argument $\pi$ would then be sent to the $\CredVerifier$ that would be reduced to check the validity of $\pi$ to verify the validity of the credential and the satisfaction of $P$.

Importantly, we note that expressing the $\BLSCES.\verify()$ predicate as an arithmetic circuit, defined over a finite field, may be quite costly. In fact, representing pairing computations as part of an arithmetic circuit requires to manipulate elements of an extension field, and, additionally, expressing the ``hash to curve'' algorithm as a set of arithmetic gates can be expensive and alter the overall performances of the protocol\footnote{The cost of representing the pairing computation as an arithmetic circuit depends on the setting of interest. In fact, we know that the existence of ``high-degree'' twists~\cite{DBLP:conf/sacrypt/BarretoLS03}, i.e. of degree $> 2$, can be used to decrease the degree of the extension field over which arithmetic is carried out to compute the pairing, see~\cite{DBLP:conf/pkc/CostelloLN10}.}.

In the following, we sketch a method to efficiently compute $\HashToGOne$ ``in the SNARK'' while moving the last step of the $\BLSCES.\verify()$ algorithm, the pairing check, ``outside of the SNARK''.
Our attention will be focused on the $\BNCurve$ pairing group as used in Ethereum~\cite{DBLP:conf/sacrypt/BarretoN05,DBLP:conf/uss/Ben-SassonCTV14}.
We recall that this curve is defined over the finite field $\FF_p$, where:
    \[
        p = \mathtt{0x30644e72e131a029b85045b68181585d97816a916871ca8d3c208c16d87cfd47}
    \]
and where the group order is:
    \[
        r = \mathtt{0x30644e72e131a029b85045b68181585d2833e84879b9709143e1f593f0000001}
    \]
These parameters have been chosen so that $r - 1$ is highly 2-adic (this is key to implement efficient SNARK-based applications which rely on radix-2 FFTs).
The embedding degree (with respect to $r$) is $k = 12$.
The curve equation is given by $y^2 = x^3 + 3$, and it admits a sextic twist\footnote{This allows to compress the representation of the elements of $\GRP_2$ by a factor of $6$ via the group isomorphism induced by the twist.}.

In the following, we denote by $\rhs$ the right hand side of the curve equation. Namely, $\rhs = x^3 + 3$.

\paragraph{Efficient representation of $\HashToGOne$ as an arithmetic circuit}


First of all, we notice that $\BNCurve$ has a \emph{cofactor} equal to one. Informally, this means that ``if we lend on the curve after hashing, then we lend in the prime order subgroup of interest'' (i.e.~we work over the whole curve).\footnote{Elliptic curves with cofactors $>1$ need to be handled with some care. In fact, in this setting, some points may satisfy the curve equation, but may not lie in the safe/secure prime order subgroup in which the discrete logarithm is hard. As a way to prevent this, one needs to ``multiply by the cofactor'' to ``move the element to the safe subgroup''.}

In order to represent a ``hash to the curve'' function as an arithmetic circuit, we propose to modify the ``try-and-increment'' algorithm~\cite[Algorithm 1]{Tibouchi2012ANO}.


By looking at the structure of the algorithm, it appears clearly that representing it as an R1CS program would not be a good fit. In fact, as the name indicates, this algorithm is probabilistic and one cannot predict with certainty if hashing the input message $\msg = (\CEAS || \credSize' || i || \cred'[i])$ will yield a point satisfying the curve equation.
As a consequence, and in order to efficiently encode the $\HashToGOne$ algorithm as an arithmetic circuit, we propose to extend the message structure, used in the $\BLSCES$ construction, to $\msg = (\CEAS || \credSize' || i || \cred'[i] || c)$, where $c$ is the counter as defined in~\cite[Algorithm 1]{Tibouchi2012ANO}. Hence, it is possible to run the ``try-and-increment'' algorithm  ``outside of the SNARK'' in order to find the counter $c$ that allows to ``lend on the curve''.
These counter values (for all signed messages) can either be adjoined by the credential issuer to each signature in the issued credential, or can be successively incremented from a known starting value in order to allow the credential holder to retrieve them during the \emph{witness generation process}.

In order to generate a zk-SNARK, $\CredHolder$ passes the message as a witness in the circuit and runs the final iteration (which is deterministic and efficient) of the ``try-and-increment'' algorithm in the SNARK. This allows to prove that the output of the hash function is a point of $\GRP_1$.

A problem remains though. In fact, we can see in the ``try-and-increment'' algorithm that the $y$-coordinate of the resulting point is retrieved by squaring the right hand side of the curve equation (i.e.~$y = \sqrt{\rhs},\ \text{where}\ \rhs = x^3 + 3$).
While we can use algorithms such as \emph{Tonelli-Shanks}~\cite{MR0371855,Tonelli} to compute the square root in the Galois field, we propose to follow another approach.

Instead of computing the square root of $\rhs$, we propose to verifiably prove that such a $y$-coordinate exists. In other words, we prove that $\rhs$ is a quadratic residue in the field.
This can be done using \emph{Euler's criterion}~\cite{euler1,euler2}, by proving that $\rhs^{(p-1)/2} \equiv 1 \mod p$.\footnote{We note however, that emulating arithmetic in the base field in the circuit (defined over the scalar field) induces some overhead.} This, in fact, is equivalent to show that $\exists\ q \in \FF_p\ \text{s.t.}\ \rhs^{(p-1)/2} - 1 = q * p$, and we know that such equality can efficiently be enforced as a set of arithmetic gates provided $q$ is given as input to the circuit (as part of the circuit witness). Moreover, computing $\rhs^{(p-1)/2}$ can be done by using an efficient exponentiation algorithm, such as the ``square-and-multiply'' approach~\cite[Chapter 14]{HAC}, which allows to compute $\rhs^{(p-1)/2}$ in $\log_2((p-1)/2)$. This can be represented efficiently as a set of arithmetic gates since $p$ is encoded on $254$ bits.

Once the $\HashToGOne$ function has been evaluated ``in the SNARK'', the $x$-coordinate used to compute $\rhs$ is added to the public inputs and sent to the verifier. Nevertheless, we know that both points $Q = (x, y)$, and $-Q = (x, -y)$ satisfy the curve equation. As such, sending only $x$ to the verifier will not be sufficient to retrieve the full point\footnote{Since, we know that given a $x$ coordinate, only two points can be retrieved, it is sometimes sufficient to assume that the party receiving $x$ will try both points $(x, y)$ and $(x, -y)$}.
However, we note that, the ``try-and-increment'' algorithm internally uses a secure hash function $H$, such as $\algostyle{sha256}$, that instantiates a Random Oracle (RO). Such function may produce digests which are longer than the binary encoding of elements of the base field. For instance, $\algostyle{sha256}$ outputs $256$-bit long digests, while the elements of $\FF_p$ are encoded on $\log_2{(p)} = 254$ bits.
In such circumstances, we can follow the method described in~\cite{Tibouchi2012ANO} and use the remaining bits of $H(m)$, i.e.~the last $|H(m)| - log_2(p)$ bits of the digest $H(m)$, in order to represent a power of $(-1)$, such that the $y$-coordinate is obtained as $y = (-1)^\variable{exp} \sqrt{\rhs}$, where $\variable{exp} = \sum_{i=0}^{|H(m)| - log_2(p)} 2^i$. As such, the remaining bits of the $H(m)$ digest can be used to retrieve the full point representation. This technique is similar to what is usually used to compress and decompress points on an elliptic curve.

\begin{remark}
    We note that recent work on ``algebraic hash functions'' (e.g.~\cite{DBLP:conf/asiacrypt/Albrecht0R0T16}\cite{DBLP:journals/iacr/GrassiKKRRS19}) lead to hash functions defined over prime fields ($\FF_p, p>2$) which have a more efficient arithmetic circuit representation than more standard ``bit twiddling'' hash functions like $\algostyle{sha256}$~\cite{10.5555/2408113} or $\algostyle{blake2}$~\cite{DBLP:conf/acns/AumassonNWW13}. Nevertheless, while such functions could be used in order to instantiate the RO needed to implement $\HashToGOne$, it is still too soon to know whether such functions are sufficiently secure to be used as plugin replacement for a random oracle.
\end{remark}

Finally, after all the steps aforementioned have been carried out, the verifier receives the SNARK $\pi$ and $(x, \variable{exp})$ allowing to verify that the $\HashToGOne$ function has correctly been evaluated on secret input $\msg = (\CEAS || \credSize' || i || \cred'[i] || c)$, and that $\HashToGOne(\msg) = (x, \variable{exp})$. The verifier parses the $x$-coordinate as an element of $\FF_p$, and ``decompresses the point'' by computing the $y$-coordinate $y = (-1)^\variable{exp} \sqrt{x^3 + 3}$.
Then, the verifier has access to the full group element representation that can be then used in the pairing check that verifies the extracted signature $\sigExt$.

We summarize this section in~\cref{fig:zkblsces-holder} and~\cref{fig:zkblsces-verifier}.

\newcommand{\ProveExtraction}{\alg{ProveExtraction}}
\newcommand{\HashToCurve}{\alg{HashToCurve}}
\newcommand{\sha}{\alg{sha256}}

\newcommand{\SnarkStatement}{\alg{SnarkStatement}}
\newcommand{\SnarkVerifier}{\alg{SnarkVerifier}}
\newcommand{\ComputeExtractionSet}{\alg{ComputeExtractionSet}}
\newcommand{\ComputeCredLen}{\alg{ComputeCredLen}}
\newcommand{\ComputeFinalTryAndIncrement}{\alg{ComputeFinalTryAndIncrement}}
\newcommand{\ComputeEulerCriterion}{\alg{ComputeEulerCriterion}}
\newcommand{\CustomPredicate}{\alg{CustomPredicate}}
\newcommand{\Prove}{\alg{Prove}}
\newcommand{\Setup}{\alg{Setup}}
\newcommand{\srs}{\variable{srs}}

\newcommand{\eci}{\variable{ec_i}}
\newcommand{\eexp}{\variable{exp}}
\begin{figure}[h]
    \begin{pcvstack}[center]
        \procedure[linenumbering, syntaxhighlight=auto]{$\ZKBLSCES.\Setup(\secparam)$}{%
            (\sk, \pk) \gets~\BLSCES.\kgen(\secparam) \\
            \srs \gets~\SNARK.\Setup(\secparam) \\
            \pcreturn (\pk, \srs)
        }

        \pcvspace

        \procedure[linenumbering, syntaxhighlight=auto]{$\ZKBLSCES.\ProveExtraction(\srs, \cred, \CEAS, \extractionSet)$}{%
            \credSize \gets \cardinality{\cred} \\
            \pcfor i \in \extractionSet \pcdo \\
            \pcind ((x_i, \eexp_i), c_i, q_i) \gets \HashToCurve(i, \cred[i], \credSize, \CEAS) \\ 
            \pcendfor \\
            \inp \gets (\{(x_i, \eexp_i)\}_{i \in \extractionSet}, \CEAS, \extractionSet) \\
            \pi \gets \SNARK.\Prove(\srs, \ZKBLSCES.\SnarkStatement(\cred, \CEAS, \{((x_i, \eexp_i), c_i, q_i)\}_{i \in \extractionSet}, \extractionSet)) \\
            \pcreturn (\pi, \inp)
        }

        \pcvspace

    \procedure[linenumbering, syntaxhighlight=auto]{$\ZKBLSCES.\HashToCurve(i, \cred[i], \credSize, \CEAS)$}{%
        \pccomment{$c_i$ is a $\log_2(k)$-bit bit string, where $k$ is set in the protocol setup}\\
        c_i \gets 0 \\
        \pcdo \\
        \pcind \pccomment{We interpret the output of the hash function as two field elements as}\\
        \pcind \pccomment{we assume that the co-domain of the hash function is bigger than the field}\\
        \pcind \pccomment{characteristic (see section above)}\\
        \pcind (x_i, \eexp_i) \gets \sha(\CEAS || \credSize || i || \cred[i] || c_i) \\ 
        \pcind \pccomment{Compute Euler's criterion to test for quadratic residuity} \\
        \pcind \eci \gets (x_i^3 + 3)^{(p-1)/2} \\ 
        \pcind \pcif \eci \equiv 1 \mod p \pcthen \\
        \pcind \pcind q_i \gets (\eci - 1)/p \\
        \pcind \pcind \pcreturn ((x_i, \eexp_i), c_i, q_i) \\
        \pcind \pcelse\\
        \pcind \pcind c_i \gets c_i + 1\\
        \pcind \pcendif \\
        \pcwhile c_i < k \\
        \pcreturn \perp
    }
\end{pcvstack}
    \caption{First part of the algorithms added to form the $\ZKBLSCES$ protocol.}
    \label{fig:zkblsces-holder}
\end{figure}

\begin{figure}[h]
    \begin{pcvstack}[center]
        \procedure[linenumbering, syntaxhighlight=auto]{$\ZKBLSCES.\SnarkStatement(\cred, \CEAS, \{((x_i, \eexp_i), c_i, q_i)\}, \extractionSet)$}{
            \pccomment{We represent here the statement proven by the credential holder.} \\
            \credSize \gets \cardinality{\cred} \\
            \pcfor i \in \extractionSet \pcdo \\
            \pcind (x_i, \eexp_i) = \sha(\CEAS || \credSize || i || \cred[i] || c_i) \\ 
            \pcind \eci = (x_i^3 + 3)^{(p-1)/2} \\ 
            \pcind \eci - 1 = q_i * p \\
            \pcendfor \\
            \extractionSet \in \CEAS\\
            \CustomPredicate(\cred) = 1 
        }

    \pcvspace

        \procedure[linenumbering, syntaxhighlight=auto]{$\ZKBLSCES.\verify(\srs, \pk, \sigExt, \pi, \inp)$}{%
            \extractionSet' \gets \inp.\extractionSet \\
            \pcfor i \in \extractionSet' \pcdo \\
            \pcind \hgen'[i] \gets (\inp.x[i], (-1)^{\inp.\eexp[i]}\sqrt{\inp.x[i]^3 + 3}) \\
            \pcendfor \\
            b_1 \gets \extractionSet' \stackrel{?}{\in} \inp.\CEAS\\
            b_2 \gets \pair(\sigExt.\sigma, \ggen_2) \isEq \prod_{i \in \extractionSet'} \pair(\hgen'[i], \pk)\\
            b_3 \gets \SNARK.\verify(\srs, \pi, \inp)\\
            \pcreturn b_1\ \land\ b_2\ \land\ b_3
        }
\end{pcvstack}
    \caption{Final part of the algorithms added to form the $\ZKBLSCES$ protocol.}
    \label{fig:zkblsces-verifier}
\end{figure}

\section{Use cases and Conclusion}

We noticed that the ability to aggregate BLS signatures coupled with the document model described in~\cite{DBLP:conf/icisc/SteinfeldBZ01} lead to a simple and efficient construction of Content Extraction Signatures. In addition of being highly regarded for its aggregation properties and determinism, we show that the BLS signature is also of high interest in the context of digital credential issuance and identity systems design.
Furthermore, one could consider using the construction detailed in this note along existing popular tools for decentralized identity, like DIDs for instance. In this setting, the BLS-CES above could provide selective disclosure for the claims composing the DID Document that the holder's DID resolves to.

\section{Acknowledgement}

We thank Ron Steinfeld for sharing a full version of~\cite{DBLP:conf/icisc/SteinfeldBZ01}, and Kristian McDonald for interesting discussions.

\bibliographystyle{alpha}
\bibliography{references}

\end{document}